\documentclass[12pt,preprint]{emulateapj}
\usepackage{amsmath, natbib, placeins}
\begin{document}


\title{The Morphology of Hadronic Emission Models for the\\ Gamma Ray Source at the Galactic Center}
\author{Tim Linden$^{1}$, Elizabeth Lovegrove$^{2}$, and Stefano Profumo$^{1,3}$}
\affil{$^1$ Department of Physics, University of California, Santa Cruz, 1156 High Street, Santa Cruz, CA, 95064}
\affil{$^2$ Department of Astronomy and Astrophysics, University of California, Santa Cruz, 1156 High Street, Santa Cruz, CA 95064}
\affil{$^3$ Santa Cruz Institute for Particle Physics, University of California, Santa Cruz, 1156 High Street, Santa Cruz, CA, 95064}
\shortauthors{}

\begin{abstract}
Recently, detections of a high-energy $\gamma$-ray source at the position of the Galactic center have been reported by multiple gamma-ray telescopes, spanning the energy range between 100 MeV and 100 TeV. Analysis of these signals strongly suggests the TeV emission to have a morphology consistent with a point source up to the angular resolution of the H.E.S.S. telescope (approximately 3~pc), while the point-source nature of the GeV emission is currently unsettled, with indications that it may be spatially extended. In the case that the emission is hadronic and in a steady state, we show that the expected $\gamma$-ray morphology is dominated by the distribution of target gas, rather than by details of cosmic-ray injection and propagation. Specifically, we expect a significant portion of hadronic emission to coincide with the position of the circum-nuclear ring, which resides between 1-3~pc from the Galactic center. We note that the upcoming Cherenkov Telescope Array (CTA) will be able to observe conclusive correlations between the morphology of the TeV $\gamma$-ray source and the observed gas density, convincingly confirming or ruling out a hadronic origin for the $\gamma$-ray emission.  
\end{abstract}

\section{Introduction}
\label{sec:introduction}

Since COS-B and EGRET first observed a bright $\gamma$-ray source spatially coincident with the Galactic center (GC)~\citep{1985A&A...143..267B, egret_gc_source}, subsequent observations by the Large Area Telescope onboard the Fermi space observatory (Fermi-LAT) and by ground-based (imaging Atmospheric Cherenkov Telescope, ACT) $\gamma$-ray telescopes have repeatedly observed the GC region at energies spanning 100~MeV to 100~TeV. Unlike both radio and X-Ray observations of the GC, no variability has been observed in the high-energy regime, potentially indicating that the $\gamma$-ray emission mechanism differs substantially from the low energy regime~\citep{chernyakova, aharonian_novariability}. This distinction is especially stark in light of the order of magnitude increase in X-ray activity from the region observed in 2005, which was undetected in $\gamma$-ray data \citep{aharonian_novariability_xray}. This might be explained by models which generate the low-frequency (X-ray, IR, radio) emission very close to the central black hole (BH), while $\gamma$ rays are produced farther away from the BH by high-energy protons inelastically scattering off of the interstellar medium, a framework originally posited by \citet{2005Ap&SS.300..255A}, and later by \citet{2006ApJ...647.1099L,2006ApJ...648.1020L}.

The strongest limits on the morphology of the gamma-ray source HESS J1745-290 are provided by \citet{2010MNRAS.402.1877A} which used optical cameras mounted on each H.E.S.S. dish to calibrate the high-energy array, reducing the systematic error in the spatial pointing of the telescope. Using this technique, \citet{2010MNRAS.402.1877A} found the source HESS J1745-290 to be spatially coincident with the radio source Sgr A*, within a total error radius of only 13\",  excluding the supernova remnant Sgr A East as the dominant source of $\gamma$-ray emission. However, the pulsar wind nebula G359.95-0.04 cannot be ruled out as a possible source of the high energy emission, and it may play a secondary or even a dominant role in the $\gamma$-ray emission observed from the galactic center. We note, however, that this possibility would not affect the analysis and conclusions presented here if the gamma-ray emission from the PSR was produced via the injection of high-energy protons into the galactic medium. HESS J1745-290 is found to be consistent with a point source to within approximately 1.2' at the 95\% confidence level\footnote{This corresponds to 2.96 pc under the assumption of a solar position 8.5 kpc from the GC.}. While this point source resides within a diffuse $\gamma$-ray background which is itself centered around Sgr A*, the total residual flux from the inner 0.1$^\circ$ is constrained to be less than 15\% of the point-source flux~\citep{2006PhRvL..97v1102A}. Work by \citet{aharonian_novariability} further constrained this diffuse emission to be relatively independent of the GC distance - indicating it may stem from cosmic-ray background events rather than a diffuse signal corresponding to the GC. 

The LAT has also observed a point source, 1FGL J1745.6-2900c, spatially coincident with the GC region~\citep{first_fermi_catalog}, although a conclusive identification with Sgr A* was impossible due to the relatively low angular resolution of the instrument ($\sim$ 0.2$^\circ$ at 10~GeV). There are indications, however, that the GC signal observed by the Fermi-LAT is difficult to model as a simple point source.  While independent point-source analyses by both \citet{2011PhLB..705..165B} and \citet{chernyakova} produced best-fit models assuming a point-source emission, the analysis of \citet{hoopergc} found that the  $\gamma$-ray emission extends spherically out to approximately 50~pc from the position of Sgr A*, falling off with a power-law of approximately r$^{-2.6}$. \citet{hoopergc2} pointed out two potential issues with GC point-source models at Fermi energies. First, a point-source best fit systematically over-estimates the source flux in the presence of an appreciable diffuse background. Secondly, even after accounting for surrounding point-sources and Galactic plane emission, the excess emission observed in the GC and not associated to the (best-fitted) point source, exceeds the point source luminosity by approximately a factor of three - in stark contrast to H.E.S.S observations at TeV energies. This second finding is consistent with the observation by \citet{2011PhLB..705..165B} that the log-likelihood of the fit increased by 25 with the addition of a spherical symmetric parameter describing the spatial extension. 

In addition to these morphological inconsistencies, the origin of the GC source spectrum at high-energy $\gamma$-ray frequencies is puzzling. A relatively hard spectrum in the 0.1-1 GeV range significantly softens in the 1-100 GeV region, then hardens again at TeV energies, before cutting off above 10~TeV. Leptonic models have been proposed to explain both the TeV emission \citep{2004ApJ...617L.123A, 2007ApJ...657..302H} and the GeV emission \citep{2012arXiv1201.5438K}. However, no theoretically compelling scenarios have been proposed to explain the entire gamma-ray spectrum a single leptonic source population.

\citet{2007ApJ...657L..13B} examined the possibility that the TeV signal observed by HESS could be explained by a significant emission of hard protons from Sgr A*, producing the gamma-ray emission through interactions with gas in the galactic center region, similar to the earlier models of \citet{2005Ap&SS.300..255A}. In order to examine the possibility that this emission be confined to the region near Sgr A*, they employed models for the morphology of Hydrogen gas in the GC from ~\citet{2004ApJ...604..662R}, which include a large ``ring" of overdense gas surrounding the GC at distances from 1-3~pc. This dense structure, aptly named the circumnuclear ring (hereafter CNR), has long been observed by far-infrared instruments as a torus-shaped structure inclined 20$^\circ$ with respect to the Galactic plane and surrounding a relative underdensity of gas within the inner pc around the GC~\citep[see e.g.][]{1982ApJ...258..135B}.  \citet{2007ApJ...657L..13B} then calculated the diffusion of charged protons within tangled magnetic fields with strength proportional to the local Hydrogen density. Using these models, they calculated the expected interaction probability between high energy protons and the CNR as a function of energy, finding approximately 73\% of protons emitted from the GC to encounter the CNR for protons with energies between 1-2.5~TeV. However they found this interaction probability to fall quickly with increasing proton energy due to the increasing gyroradius of high energy protons - only 47\% and 5\% of protons are found to encounter the CNR at energies of 10 and 100 TeV respectively. Due to the extreme overdensity of hydrogen gas in the CNR, \citet{2007ApJ...657L..13B} assumed that protons encountering the CNR would lose energy quickly through pion collisions, while protons which avoid the ring would lose energy at larger radii, producing an extended gamma-ray emission at higher energy.

Recently \citet{chernyakova} further examined the hadronic scenario, additionally positing that the whole GeV-TeV $\gamma$-ray emission may be explained by the injection of high energy protons at the GC. This interpretation has several natural advantages in explaining the entire $\gamma$-ray spectrum, including: (1) the hard spectrum below 1~GeV is naturally explained by the inability of protons with kinetic energy below $\sim$300~MeV to produce $\pi^0$ in p-p collisions, (2) the bump in low energy $\gamma$-ray emission is produced by diffusively trapped protons which lose significant energy while propagating through the GC region, (3) the flat spectrum at TeV energies is explained by protons which propagate rectilinearly through the GC region, without losing significant energy in p-p collisions, which provides a convincing match to the $\sim$E$^{-2}$ $\gamma$-ray spectrum by using the E$^{-2}$ proton injection spectrum derived from first-order Fermi acceleration, and (4) the turnover between the GeV and TeV emission is naturally explained as the transition between diffusive and rectilinear motion - creating a soft spectrum between these two energy regimes. \citet{chernyakova} also consider the possibility that two distinct proton populations are responsible for the low- and the high-energy $\gamma$-ray emission. 

In the present study we focus on the {\em morphology} of the GC $\gamma$-ray source. Specifically, we introduce a detailed model of the interstellar gas distribution near the GC, and demonstrate that the $\gamma$-ray morphology from hadronic emission is determined primarily by the distribution of target gas rather than by the parameters describing the diffusion of high-energy protons. We find that in any scenario for cosmic ray diffusion, the bulk of the high-energy emission falls within the point-spread function of all current $\gamma$-ray telescopes. However, we note that this will not be the case for the proposed Cherenkov Telescope Array (CTA), which will have the angular resolution required to observe a morphology in the $\gamma$-ray source which shadows the observed gas density \citep{CTAref}. The outline of this study is as follows: Sec.~\ref{sec:gas} describes the model we employ for the GC gas density; Sec.~\ref{sec:qualitative} gives qualitative arguments to describe cosmic ray proton propagation in the GC region, while in Sec.~\ref{sec:prop} and \ref{sec:results} we present our results for the proton population and for the $\gamma$-ray emission in hadronic models for the GeV-TeV emission.

\section{Gas Density near the Galactic Center}
\label{sec:gas}

\begin{figure}
                \plotone{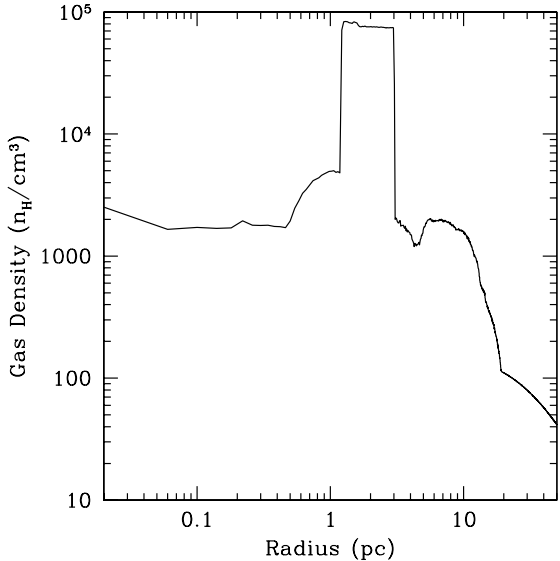}
                \caption{Gas density (n$_H$/cm$^3$) averaged over solid angle as a function of radius from the GC (pc) obtained from a combination of the work of \citet{2007A&A...467..611F} and \citet{2012arXiv1201.6031F} (see the text). The major feature in the gas density stems from the circum-nuclear ring (CNR, 1.2~pc to 3.0~pc) which contributes a gas density nearly two orders of magnitude larger than any other structure.}
                \label{fig:gasdensity}
\end{figure}

Detailed knowledge of the gas density is critical to accurately describing hadronic $\gamma$-ray emission from the GC region. Here we employ the model of ~\citet{2012arXiv1201.6031F}, valid for the inner $\sim$10~pc around the GC. This model not only includes a diffuse halo, but also contributions from the CNR, Sgr A East, M-0.13-0.08 and M-0.02-0.07 as well as bridges and streamers connecting them. We make several necessary simplifications to the model before utilizing it within our numerical code. First, we create a spherical model by calculating the average gas density at each radius $r$. In doing this, we take a central value for the position of M-0.13-0.08 and make geometric approximations for the position of the various streamers, keeping the total volume consistent with~\citet{2012arXiv1201.6031F}. We note that this will have the effect of artificially making the $\gamma$-ray emission spherically symmetric. However, in the limit that both the proton diffusion is spherically symmetric, and the probability of multiple interactions between a single cosmic ray proton and the interstellar medium is low, the total emission from within a given radial bin is conserved when the gas density is smeared. We will discuss these assumptions in detail later, and we will show that they are valid for the majority of the parameter space discussed. Secondly, we ignore the thermal distribution of gas, which is valid in the limit where we only consider collisions with highly-relativistic protons. 

To extend our simulations beyond 10~pc from the GC in order to capture the region relevant for the Fermi-LAT PSF, we adopt the model of Eqs. 18 and 19 of~\citet{2007A&A...467..611F} and impose again spherical symmetry. With these assumptions, we obtain gas densities which are lower than those in \citet{2012arXiv1201.6031F} by an order of magnitude at a radius of 10~pc. Since the models of \citet{2012arXiv1201.6031F} are only suffering incompleteness at radii greater than 10~pc (due to unmodeled sources), we assume the gas density at a given radius to be the larger of the values quoted by \citet{2007A&A...467..611F} or \citet{2012arXiv1201.6031F}. This causes us to switch between models at a radius of 19.2~pc. In Figure~\ref{fig:gasdensity} we show the resulting density of hydrogen gas in our simulation as a function of the radius from the GC. We note that the morphology is dominated by the CNR, which provides a large boost to the gas density between 1-3~pc.

\section{A Qualitative Model}
\label{sec:qualitative}

\begin{deluxetable*}{ccc}
\tabletypesize{\scriptsize}
\tablewidth{0pt}
\tablehead{
\colhead{Parameter} & \colhead{\citet{chernyakova} Value} & \colhead{Value Adopted Here}}
\startdata
Radial Size of Simulation Region & 3 pc & 50 pc \\
Duration of Proton Emission & 300 - 10$^4$ yr & 10$^{10}$~yr\\
Proton Injection Spectrum & -1.9 -- -2.0 & -1.9 \\
Hydrogen Density during proton diffusion & 1000 cm$^{-3}$ & 1000 cm$^{-3}$ \\
Hydrogen Density for $\gamma$-ray calculation & 1000 cm$^{-3}$ & See Figure 1 \\
\enddata
\tablenotetext{}{\label{tab:chern} List of all input parameters which differ between the assumptions of \citet{chernyakova} and the present work. }
\end{deluxetable*}

In order to understand the morphology of the hadronic emission, we consider propagation in four limiting regimes, controlled by two parameters. Protons may propagate either rectilinearly or diffusively through the GC region, and they may either undergo many, or much less than one, collision with the surrounding gas. While intermediate cases are possible, we find these four limiting cases to confine the expected proton morphology. 

We first investigate the propagation of protons in the rectilinear regime. This can be thought of as diffusion with a mean-free-path exceeding the confinement region of the simulation, yielding a diffusion constant D~=~$l_{sim}^2/{(6\tau)}$, where $l_{sim}$ is the region we are considering, and $\tau$ is the propagation time out of the region. For our 50~pc simulation and assuming relativistic propagation velocities, this corresponds to a minimum D $>$~7.7~x~10$^{29}$~cm$^2$s$^{-1}$ for particles to propagate rectilinearly out of the simulation zone. We can calculate the probability of a proton-proton interaction between the cosmic ray and a target gas molecule as 

\begin{equation}
P_{pp} =~{\sigma_{pp}(E)\int_{\vec r(0)}^{\vec r(t)}\rho_{H}(\vec r(t'))dt'}
\end{equation}

where $\sigma_{pp}(E)$ is the cross-section between a proton with Energy E and a cold target proton, $\rho_{H}$ is the gas density as a function of position, and $\vec r(t)$ is the radial position as a function of time from the injection of the cosmic ray until it leaves the diffusion zone. In the case of spherical symmetry and rectilinear propagation out of the center of the diffusion zone, we set r(t')~=~ct', r(0)~=~0, and r(t)~=~r$_{max}$, and then using the gas density shown in Figure~\ref{fig:gasdensity}, infer an interaction probability of $\sim$2.5\% between a given proton and the target gas before the proton escapes the 50~pc region. In this case, relativistic protons lose little energy as they propagate through the region, and the proton spectrum mirrors the injected spectrum. Since the number of protons in a spherical shell $\delta$r is constant for rectilinear propagation, the resulting $\gamma$-ray morphology mirrors the gas distribution shown in Figure~\ref{fig:gasdensity}. In this scenario 87\% of the total $\gamma$-ray emission is concentrated within the inner 3~pc surrounding the GC. If, as expected, this scenario describes the propagation of the $\sim$10~TeV protons responsible for the TeV $\gamma$-ray signal, then this ratio stands in excellent agreement with HESS observations indicating at least 85\% of $\gamma$-ray TeV emission to be contained in the inner 3~pc. We note that so long as we assume rectilinear propagation, this conclusion still holds in the case of a true 3D model of the gas density, and the final emission morphology is calculable by convolving the $\rho(r) \sim r^{-2}$ density of protons with the observed gas density  (e.g., \citet{2012arXiv1201.6031F}) and integrating over the line of sight. 

We note that for the gas densities and distance scales modeled here, the second possible regime, where the diffusion is rectilinear but protons undergo more than one collision with the target gas is insignificant, since the probability of having even one collision is only $\sim$2.5\%. Thus we exclude the second regime. 

A third regime exists when protons propagate diffusively through the region, but interact with the Galactic gas fewer than once. In this case, we can assume that the energy spectrum of protons remains constant with radius, and calculate the density of protons to fall off as $\rho(r) \sim r^{-1}$, causing the number of protons within a radial bin to increase linearly with radius. In this case, we can directly calculate the morphology of the resulting $\gamma$-ray emission by using the gas density shown in Figure~\ref{fig:gasdensity}, but weighting the gas density by the number of protons in each radial bin, which varies linearly with r. Weighted in this manner, we calculate the average n$_H$ density as 410 cm$^{-3}$, and find the minimum diffusion constant for which less than 0.1 interaction occurs to be 2.4~x~10$^{27}$~cm$^{3}$s$^{-1}$. In this limit, the expected $\gamma$-ray morphology results from convolving the gas density in Figure~\ref{fig:gasdensity} with a proton spectrum which is constant in radius and has an overall density $\rho(r) \sim r^{-1}$: the $\gamma$-ray morphology is therefore still dominated by the CNR, mirroring what was found in first regime.

Lastly, we evaluate the regime where diffusive cosmic rays are expected to interact multiple times with interstellar gas before diffusing out of the GC region. In this case, the final cosmic ray energy spectrum is not radius independent, and cannot be easily computed. Furthermore, we note that in this regime, the non-spherically symmetric nature of the gas becomes important in determining the steady-state cosmic-ray density, as interactions which occur while protons move through particularly high-density regions may have an effect on the surrounding proton distribution. However, we note that, so long as the injection proton energy spectrum is sufficiently steep that partially cooled high energy protons are subdominant to the injected low energy population, this cosmic ray density is unambiguously constrained to fall off faster with increasing radius than in the one-interaction regime above. In order to provide more quantitative calculations, we defer to a numerical model for the propagation of high energy particles throughout the GC region. 

We note that the r$^{-1}$ dependence of cosmic-rays propagating diffusively out of the galactic center, closely mirrors the brightness profile of gamma-ray emission calculated by~\citet{chernyakova} (Figure 7) for cosmic rays of 1-10 GeV. This is expected, as these models use a constant Hydrogen density, creating a one-to-one correspondence between the proton density and the brightness profile. Notice that the diffusion constant at 1-10 GeV employed by \citet{chernyakova} fall between 9.5~x~10$^{26}$ and 7.6~x~10$^{27}$ cm$^2$s$^{-1}$, which lands squarely in the regime of diffusion with fewer than one interaction per cosmic ray (using a region with radius 3~pc and a gas density of 1000 cm$^{-3}$). In the lower energy bin of 0.1-1~GeV, the significantly smaller diffusion constant leads to multiple interactions between high energy protons and the target gas, leading to a more constrained distribution, as predicted in our qualitative model.
 
\section{Propagation of High-energy Protons}\label{sec:prop}

In order to simulate the propagation of protons from a central source, we adopt the formalism of \citet{aloisioprop}, which seeks to model spherically symmetric diffusion while avoiding the issue of superluminal propagation. We note that this cosmic-ray diffusion model is nearly identical to that employed by \citet{chernyakova} and we refer the reader to that work for relevant details of our simulation parameters. In Table~\ref{tab:chern} we list all differences between our model and that of \citet{chernyakova}. Most importantly, we employ a much larger simulation region with a 50~pc radius, in order to capture the entirety of emission within the Fermi PSF. In order to obtain a steady state diffusion solution over this period, we must assume that the the emission continues for periods longer than the 10$^{4}$~yr assumed in \citet{chernyakova}. Since we are only concerned with the upper limit for the source extension in this work, we allow the emission to continue for 10$^{10}$~yr, but note that the equation becomes steady state after  approximately 10$^{6}$~yr.

We adopt a differential proton injection spectrum which follows a power law $\sim$E$^{-1.9}$ with an exponential cutoff at 100~TeV. We normalize this injection spectrum in order to obtain a $\gamma$-ray intensity matching the Fermi and HESS fluxes reported by \citet{chernyakova} at a radius of 3~pc, and then extend our simulation from this point out to 50~pc. We find that this translates to an injected proton luminosity of 8.8~x~10$^{36}$~erg~s$^{-1}$. 

For simplicity, we utilize an average target density  n$_H$~=~1000~cm$^{-3}$ in our calculation of the final proton density (but not for the final $\gamma$-ray emission, for which we employ the target gas density given by Figure~\ref{fig:gasdensity}). This is an immaterial assumption, so long as we reside in a limit where the average proton undergoes much less than one collision with gas - an assumption which holds throughout the vast majority of our parameter space. We calculate the $\gamma$-ray emission from the steady-state proton density distribution employing the formalisms of \citet{2006ApJ...647..692K}. We note that we consider here only contributions from $\gamma$-rays produced directly in the p-p collision, and ignore possible $\gamma$-ray contributions due to the inverse-Compton scattering (ICS) of photons by leptons produced in the same collisions. This approximation is reasonable both because ICS contributes at significantly lower energies than the direct $\gamma$-ray channel (and will thus be subdominant to the larger proton flux at lower energies), and because the magnetic fields are expected to peak strongly in this region, forcing the ICS to be highly subdominant to synchrotron production~\citep{2010Natur.463...65C}. We note that variations in the diffusion parameters are constrained by the transition between the GeV-TeV spectrum, and yielded no qualitative changes in the findings outlined below.

\section{Results}
\label{sec:results}

\begin{figure}
                \plotone{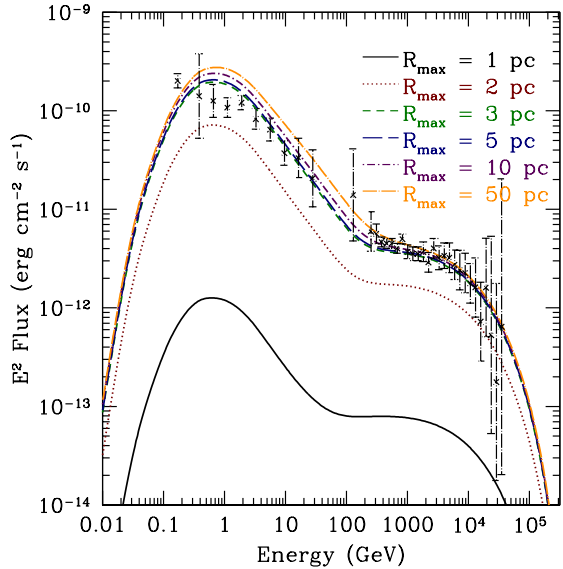}
                	\caption{\label{fig:spectrum} Spectral energy distribution of $\gamma$-ray emission within radial cones of various size ranging from r~$<$~1~pc to  r~$<$~50~pc. We note that the emission falls off very quickly after 3~pc, with almost no additional intensity for additional radial bins beyond this point. The overall intensity is normalized such that the emission within 3~pc matches the Fermi and HESS fluxes as reported by \citet{chernyakova}. Due to the higher n$_H$ in our simulation and longer emission period, this corresponds to a lower proton injected luminosity of 7.0~x~10$^{35}$~erg~s$^{-1}$ into protons.}
\end{figure}

\begin{figure}
                \plotone{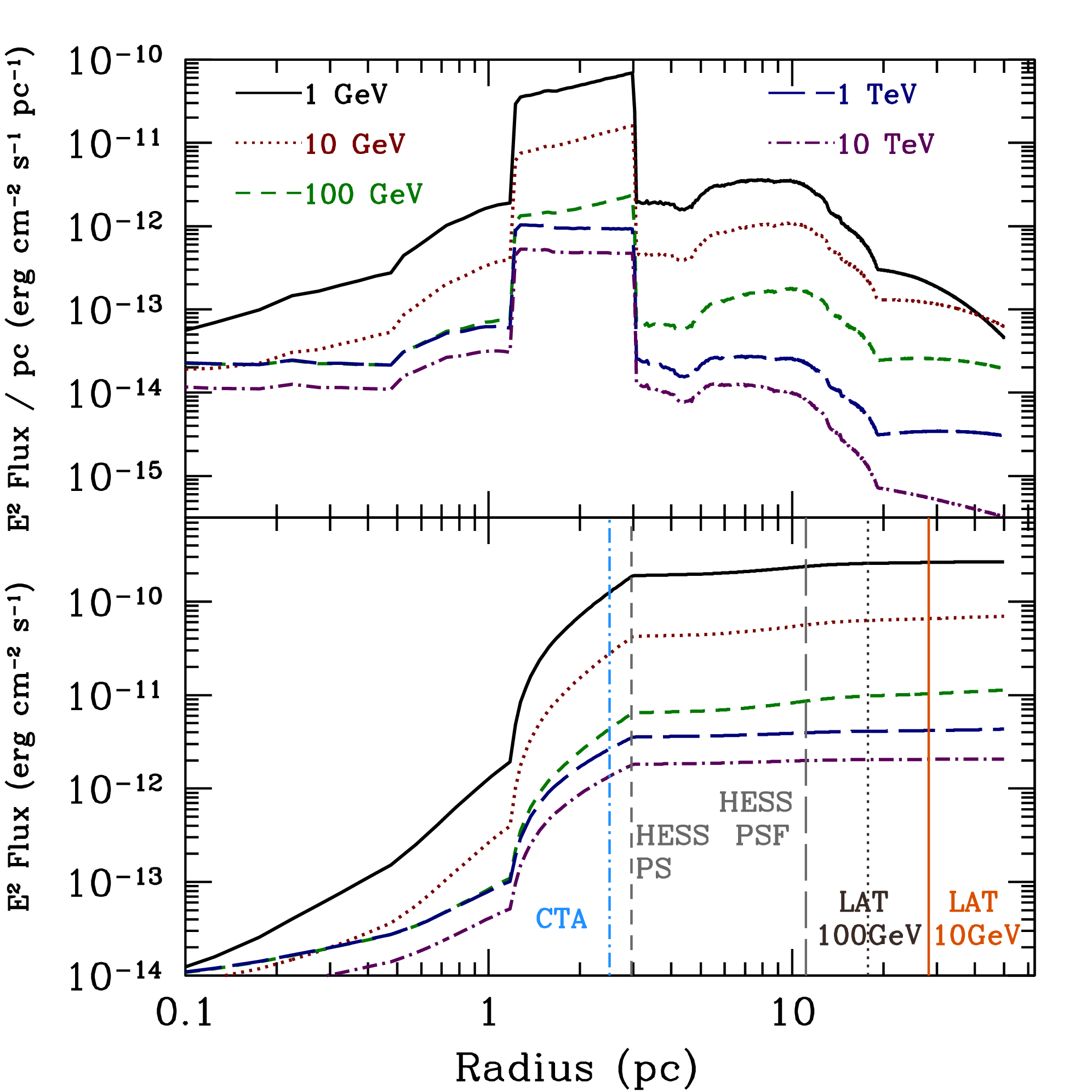}
                	\caption{ \label{fig:radial} Differential (top) and integrated (bottom) radial emission for $\gamma$-ray energies of 1~GeV (solid black), 10~GeV (dotted red), 100~GeV (green dashed), 1~TeV (blue long dash) and 10 TeV (purple dot-dashed). In the bottom panel we show vertical lines corresponding to the angular resolution of the Fermi-LAT at 100 GeV (orange solid), Fermi-LAT at 10~GeV (brown dotted), HESS (gray long dashed), CTA (projected, blue dash-dot), as well as the HESS 95\% confidence limit on the source extension (HESS PS, gray short dashed).}
\end{figure}

In Figure~\ref{fig:spectrum}, we show the spectral energy distribution of the $\gamma$-ray signal within cones corresponding to varying radii. We note that more than 50\% of the residual emission is found within 3~pc at all energies, and the majority of this emission is created between 1-3~pc from the GC, when the emitted protons interact with the dense CNR. In Figure~\ref{fig:radial}, we instead show the differential (top panel) and integrated (bottom) emission as a function of radius at various fixed energies. In the top panel, we show the total emission in radial shells of 1~pc width, while in the bottom panel we show the integrated emission within a given radius.

We note that these features correspond closely to the expectations from Section~\ref{sec:qualitative}. For instance, at energies of 10~TeV the emission is dominated by a proton population of approximately 100~TeV, which propagates rectilinearly in our simulation. We find the $\gamma$-ray emission contained within 3~pc and 10~pc to be 87\% and 95\% respectively, matching the expectations from rectilinear propagation. At energies of 10~GeV, the $\gamma$-ray signal is dominated by 100~GeV electrons which propagate diffusively but undergo Poissonian interactions. In this case we calculate the $\gamma$-ray emission contained within 3~pc and 10~pc to be 61\% and 78\% respectively, closely matching an r$^{-1}$ proton density. 

Finally, in Figure~\ref{fig:radial} (bottom) we show with vertical lines the approximate angular resolutions of front converting events from the Fermi-LAT at 10~GeV (28~pc, orange solid), and at 100~GeV (18~pc, gray dotted), HESS (11~pc, gray long-dashed \citet{2006ApJ...636..777A}), the HESS 95\% confidence limit on the maximum source extension(3~pc, gray short-dashed, \citet{2010MNRAS.402.1877A}) and the anticipated proposed resolution of CTA \citep[2.5~pc, light blue dash-dot]{CTAref}. We note that the angular resolution for the Fermi-LAT at 1~GeV is approximately 90~pc and is outside the plotted range. For all existing telescopes, the modeled radial dependence could easily be confused with a point source located at the GC. However, similarly to HESS, CTA will be able to place limits on the source extension which are a factor of few smaller than the quoted angular resolution. In this case, CTA will clearly detect structure in the $\gamma$-ray signal, including, possibly, bright emission coincident with the CNR. We expect that this signature would provide extremely strong evidence in favor or against the hadronic nature of this emission. 

\section{Discussion and Conclusions}
\label{sec:conclusions}

We have shown that hadronic emission stemming from the inner 3~pc of the Galaxy and following a power-law injection spectrum compatible with Fermi acceleration will naturally produce an emission spectrum comparable to that observed by both HESS and the Fermi-LAT. We showed analytically and numerically that the morphology of $\gamma$-ray emission is determined primarily by the gas morphology, and the majority of the emission falls below the PSF of all current $\gamma$-ray telescopes. While this is in extremely good agreement with HESS observations reporting 85\% of the TeV emission to be confined within 3~pc of the GC~\citep{2006PhRvL..97v1102A}, the hadronic model is currently in some tension with Fermi-LAT observations which imply that the GC source may be extended. 

Specifically, our models find 71\% and 86\% of the 1~GeV $\gamma$-ray emission to fall within 3~pc and 10~pc from the GC respectively. This result stands, at face value, in some tension with Fermi-LAT observations of the GC \citep{hoopergc2}. After background subtraction, \citet{hoopergc2} find a residual emission which is not well-modeled by a point source convolved with the Fermi-LAT PSF. If modeled as an extended emission from dark matter (but independent of any dark matter properties), they derive a best-fit power-law fall-off to the emission $\sim{r}^{-2.6}$, which indicates that roughly 32\% and 53\% of the $\gamma$-ray signal should originate within 3~pc and 10~pc of the GC. 

It is important to remark that it will be difficult to definitively conclude that hadronic models are ruled out by Fermi observations, as significant background subtraction is necessary in order to determine the residual background signal which should be observed by the Fermi-LAT. This includes a dominant foreground from emission in the galactic plane which lies along the line of sight between the solar position and the galactic plane~\citep{hoopergc2}. This intensity of residual emission can be further complicated by uncertainties in the modeling of point source contributions at the galactic center~\citep{2011PhLB..705..165B}. Lastly, the $\gamma$-ray morphology for hadronic models is neither point-like nor power-law with radius, making it difficult to directly compare these results with previous works. Still, our results indicate that a substantial source extension in Fermi-LAT data is not expected from hadronic emission models.

It may appear that the conclusion above can be circumvented if the hadronic emission from the GC is highly polar, allowing protons to propagate out of the central 3~pc while avoiding the CNR. However, this is difficult to reconcile with TeV observations: with purely polar injection, we find that HESS would clearly observe an extended source above and below the Galactic plane. Our conclusions would also be affected if the GeV emission mechanism were distinct from the TeV emission -- either due to different primary proton populations or to multiple low-energy hadronic sources contributing to the GeV but not to the TeV emission. 

Notice that a time dependence in the hadronic emission could be engineered to provide an energy-dependent source extension. Further multiwavelength observations will be necessary to constrain this scenario. Another alternative pertains to the possibility that the CNR is not a stable feature of the galactic-center region, but is instead transient in nature. While studies of galactic tracers such as HCN and HCO$^+$ found extremely high gas masses of approximately 10$^6$~M$_\odot$, which would be stable against tidal disruption by the supermassive-black hole~\citep{2005ApJ...622..346C, 2009ApJ...695.1477M}, very recent observations using the GREAT telescope to perform a CO excitation analysis ~\citep{2012arXiv1203.6687R} obtain a best fit which reduces the total mass of the CNR by approximately two orders of magnitude, and thus find the CNR to be susceptible to tidal disruption, implying that the feature could be transient. We note that the density used in this paper, based on the central value reported by~\citet{2012arXiv1201.6031F} stands at 2~x~10$^5$~M$_\odot$, in between these extreme values. The implication of a transient CNR in the context of this work is difficult to positively determine, as it would depend sensitively on the effect of the CNR not only on the gas density, but on the diffusion parameters of the region as well. If the latter effect is neglected, then it is possible that the fall-in of the CNR would lead to enhanced low-energy emission, as protons at this energy have built up without losing significant energy until the CNR moves into regions with higher proton density nearer the GC.

We note that while our work considers a similar model for the gas density near the GC to that of \citet{2007ApJ...657L..13B}, we obtain an energy dependence for the gamma-ray morphology which is qualitatively different. Specifically, \citet{2007ApJ...657L..13B} calculates the probability of a cosmic ray proton entering the CNR before it diffuses out of the target region. Since this probability will decrease as the protons mean free path becomes larger, they determine that the gamma-ray morphology should become more diffuse at high energy. However, this implicitly assumes that gamma-rays moving through the CNR efficiently lose their energy before leaving the region. However, employing the model for Hydrogen in the CNR determined by~\citet{2012arXiv1201.6031F}, and using a model where the propagation of TeV photons is rectilinear on the distance scale inhabited by the CNR, we find that protons escape from the CNR with only negligible energy losses. Additionally, we find that these energy losses remain negligible even through the majority of the diffusive regime. 

This creates a significant difference in the necessary proton injection spectra necessary to match the HESS dataset in this work, compared to that of  \citet{2007ApJ...657L..13B}. In our work the existence of rectilinear proton propagation throughout the HESS regime implies that the injected E$^{-1.9}$ proton produces an almost equivalent E$^{-1.9}$ gamma-ray spectrum. However, in the work of \citet{2007ApJ...657L..13B} the decreasing interaction probability between high energy protons and the molecular gas requires a much harder injected proton spectrum (E$^{-0.75}$) to account for the observed gamma-ray spectrum. Finally, we note that while the percentage of high energy emission contained within the inner 3~pc in our model is similar to the probability of a 1 TeV proton encountering the CNR as determined by \citet{2007ApJ...657L..13B}, this is fairly coincidental: the relative emission in our model depends sensitively on the Hydrogen gas density outside of the inner 3~pc, which is not taken to account in the model of \citet{2007ApJ...657L..13B}.

Finally, we note that the greatly improved angular resolution of CTA will provide a much crisper picture than current instruments - potentially leading to the observation of both a central deficit of $\gamma$ rays in the central pc from the GC, and of a bright ring of emission coincident with the position of the CNR if the GC source is of hadronic nature. This morphology would be difficult to explain with leptonic or dark matter scenarios, which would not be expected to correlate with the local gas density. Thus we believe that future CTA observations will convincingly confirm or rule out the hadronic emission scenario presented here. \\

We are grateful to Felix Aharonian, Johann Cohen-Tanugi, Dan Hooper, Tesla Jeltema and Max Wainwright for helpful comments. This work is partly supported by NASA grant NNX11AQ10G. SP acknowledges support from an Outstanding Junior Investigator Award from the Department of Energy, and from DoE grant DE-FG02-04ER41286. \\ \newpage

\bibliography{gc} 

\end{document}